# Electrodialytic removal of tungsten and arsenic from secondary mine resources – Deep eutectic solvents enhancement

J. Almeida[1,3*], R. Craveiro[2], P. Faria[3], A. Santos Silva[4], E.P. Mateus[1], S. Barreiros[2], A. Paiva[2], A. B. Ribeiro[1*]

[1]CENSE, Department of Sciences and Environmental Engineering, School of Science and Technology, NOVA University Lisbon, Caparica Campus, 2829-516 Caparica, Portugal

[2]LAQV@REQUIMTE, Department of Chemistry, School of Science and Technology, NOVA University Lisbon, Caparica Campus, 2829-516 Caparica, Portugal

[3]CERIS and Department of Civil Engineering, School of Science and Technology, NOVA University Lisbon, Caparica Campus, 2829-516 Caparica, Portugal

[4]Department of Materials, National Laboratory for Civil Engineering, 1700-066 Lisbon, Portugal

*Corresponding author.

E-mail addresses: js.almeida@campus.fct.unl.pt (J. Almeida); abr@fct.unl.pt (A. B. Ribeiro)

HIGHLIGHTS

- Proof of concept in coupling DES + ED for As and W removal from mining residues

- Best As and W extraction achieved with ChCl:MA (1:2) and ChCl:OA (1:1)

- DES and electrical current use improved extraction for As (22 %) and W (11 %)

- From total As and W extracted, electromigration achieved 82 % for As and 77 % for W

ABSTRACT

Tungsten is a critical raw material for European and U.S. economies. Tungsten mine residues, usually considered an environmental burden due to e.g. arsenic content, are also secondary tungsten resources. The electrodialytic (ED) process and deep eutectic solvents (DES) have been successfully and independently applied for the extraction of metals from different complex environmental matrices. In this study a proof of concept demonstrates that coupling DES in a two-compartment ED set-up enhances the removal and separation of arsenic and tungsten from Panasqueira mine secondary resources. Choline chloride with malonic acid (1:2), and choline chloride with oxalic acid (1:1) were the DES that in batch extracted the average maximum contents of arsenic (16 %) and tungsten (9 %) from the residues. However, when ED was operated at a current intensity of 100 mA for 4 days, the extraction yields increased 22 % for arsenic and 11 % for tungsten, comparing to the tests with no current. From the total arsenic and tungsten extracted, 82 % and 77 % respectively were successfully removed from the matrix compartment, as they electromigrated to the anolyte compartment, from where these elements can be further separated. This achievement potentiates circular economy, as the final treated residue could be incorporated in construction materials production, mitigating current environmental problems in both mining and construction sectors.

Abbreviations

"W" – Tungsten; "As" – Arsenic; "ED" - Electrodialytic; "DES" - Deep Eutectic Solvents; "ChCl:OA" – Choline Chloride/ Oxalic Acid; "ChCl:MA" – Choline Chloride/ Malonic Acid; "ChCl:LA" – Choline Chloride/ Latic Acid; "PA:U" – Propionic Acid/ Urea

1. INTRODUCTION

Tungsten (W) is a critical raw material with a wide range of uses, being the largest in cemented carbides production, followed by mill products, alloys and steels (Cuesta-Lopez, 2017). China owns 82 % of the world's W mine production (U.S. Geological Survey, 2019). In addition, other W primary ores commercially mined, such as scheelite ($CaWO_4$) and wolframite ($(Fe,Mn)WO_4$), are becoming gradually limited (Yang et al., 2016). The European Union and the United States of America recognized W, respectively, as one of the 27 Critical Raw Materials (European Comission, 2017), and as one of the 35 mineral commodities considered critical to U.S. National Security and Economy (U.S. Geological Survey, 2018). Tungsten deposits have typically a grade of less than 1 % $WO_3$. Additionally, due to the waste rock that needs to be removed to access the ore, W mines produce volume-wise more waste than has originally been mined. The accumulation of residues in open pits generates serious landscape and other environmental problems (Schmidt et al., 2012).

Panasqueira mine (located in the *Centro* region of Portugal) has been active for more than one century and is one of the largest tin (Sn) - W deposits in Europe, with estimated 9.7 Mt of ore resources (Candeias et al., 2014). These mining residues are low grade secondary resources that contain not only W, but also other elements of environmental concern, as arsenic (As) from arsenopyrite (FeAsS) (Candeias et al., 2014), an element which toxicity strongly affects the individuals exposed (Coelho et al., 2014).

The electrodialytic (ED) process is a treatment technology, that has been studied over the last three decades, to remove heavy metals from several environmental matrices: (1) soil (Ribeiro and Mexia, 1997), (2) sludge (Guedes et al., 2015), (3) fly ash (Ferreira et al., 2005), (4) timber waste (Ribeiro et al., 2000) and (5) mine tailings (Hansen et al., 2008; 2007; Zhang et al., 2019). When a low-level direct current is applied between a pair of electrodes, the removal or separation of substances from polluted matrices is promoted (Ribeiro et al., 2016). Despite promising data, the technology readiness level (TRL) remains far from being introduced as an efficient process into the market (Lacasa et al., 2019).

Deep eutectic solvents (DES) have been considered the solvents of the XXI century, having successful results in the extraction of metals from various sources (Abbott et al., 2017; 2015; 2011; 2005; Schaeffer et al., 2018; Söldner et al., 2019; Su et al., 2018). DES are now a feasible option for scale-up purposes, since they are composed by inexpensive and abundant raw materials, and their formulations are simple and reproducible (Smith et al., 2014; Zhang et al., 2012). DES are a mixture of hydrogen bond donors and hydrogen bond acceptors. When mixed at a certain molar ratio, the melting point of the mixture becomes significantly lower than that of the original components (Abbott et al., 2003). Additionally, tuning the water content and composition of DES can change their properties, favoring their versatility for several applications. Through the use of strong acids, as sulfuric and nitric acids, the yields in metals extraction could achieve up to 97% (Gong et al., 2019; Pena-Pereira and Namieśnik, 2014; Shen et al., 2018). However, their negative impacts are also widely recognized (Shen et al., 2019). The use of natural DES offers advantages in terms of costs, toxicity and biodegradability (Pena-Pereira and Namieśnik, 2014).

To the best of our knowledge, the use of DES to enhance the ED process removal/separation yield has not been reported. Mining residues can be treated with ED technologies via stirred suspensions mixtures (Zhang et al., 2019). In these cases, critical raw materials recovery and harmful compounds/elements removal may be extremely attractive. The decrease of primary sources needs and toxic risks in an industrial scale application can be reduced coupling ED and DES.

If a successful recovery of the critical raw material W and the removal of As from secondary mine resources is achieved, it will contribute to close loops, in a circular economy perspective. Secondary resources re-integration provides a platform for the study of new concepts and application of technologies, business models, and policy for sustainable circular economy approaches in several industrial sectors (Velenturf et al., 2019). Furthermore, alternative strategies to recover secondary resources in EU and US will decrease their resources dependency from other countries.

Additionally, in the construction sector, innovative alternatives for materials under basic conditions are now increasing attention since they are less prone to leach hazardous elements and their durability may not be conditioned by physical, chemical and microstructural properties changes. This opens the possibility for treated mining residues with ED and DES be further reused in construction materials, alleviating the need of primary resources.

This work aimed to: (1) assess efficient DES for W and As extraction, (2) combine the use of DES in the ED process, and (3) perform a proof of concept of the feasibility to couple ED treatment and DES to separate W and As from the matrix, into a different compartment.

## 2. Materials and Methods

### 2.1 Materials

Mining residues were collected at Panasqueira mine, Covilhã, Portugal (40°10'11.0604"N, 7°45'23.8752"W). The plant produces ~900 t $WO_3$/year, and the collected residues corresponds to the rejected fraction from the sludge circuit, directly pumped into a dam. The collected mining residues were filtered by vacuum with filter paper 42 (Whatman,

Germany) and left to dry for 48 h at room temperature. All the experiments were conducted with filtered and dried residues.

Deep eutectic solvents were prepared with choline chloride (ChCl) CAS 67-48-1 (≥ 99 %, Sigma-Aldrich, Germany), malonic acid (MA) CAS 141-82-2 (99 %, Sigma-Aldrich, Germany), oxalic acid (OA) CAS 144-62-7 (≥ 99 %, Sigma-Aldrich, Germany), propionic acid (PA) CAS 79-09-4 (99 %, Sigma-Aldrich, Germany), urea (U) CAS 57-13-6 (≥ 99.5 %, Merck, Germany) and DL-lactic acid (LA) CAS 50-21-5 (≥ 85 %, Tokyo Chemical Industry, Japan). For water content measurements, Hydranal Coulomat AG CAS 67-56-1 (Honeywell, Germany) was used.

### 2.2 Experimental set-up

Twelve experiments were carried out in duplicate, according to the conditions presented in Table 1. Due to the heterogeneity of the sample, high standard deviations on the results may be expected. A liquid/solid (L/S) ratio of 9 was selected for the experiments, since ratios between 7 and 12 allow the suspension of the sample, improving extraction efficiencies (Nystroem et al., 2005).

Experiments were designed according to the following objectives:

1) To access the best DES combination for As and W extractions
    a. Experiments: 1-4
    b. Mixture: 2.2 g of mining residues with 20 mL of DES

2) To study the influence of the current intensity with DES on As and W extractions
    a. Experiments: 5-8
    b. Mixture: 11.1 g of mining residues within 98.5 mL of deionized water and 1.5 mL of DES

    c. Set-up: a power supply was connected to a pair of electrodes, and the cathode and the anode ends were placed inside a beaker, in order to generate an electric field.

    d. Current intensities: 50 and 100 mA

3) To test the feasibility of coupling DES with the ED process in order to increase As and W extraction and allow their later separation in a different compartment.

    a. Experiments: 9-12

    b. Mixture (cathode compartment): 39 g of mining residues (1.18 g/cm$^3$) with 345 mL of deionized water and 5 mL of DES

    c. Anode compartment: 250 mL of 0.01M $NaNO_3$ (anolyte)

    d. ED set-up: two-compartment cell, as seen in Figure 1. A

The level of the two compartments in the ED set-up is different. The anode compartment is full, and the cathode compartment was filled up to ~400 mL out of a 500 mL capacity (not completely so that no overflow would occur during the stirring), which may potentiate osmosis phenomena from the anode to the cathode compartment during the treatment.

The two-compartment ED set-up (Figure 1) was an acryl XT cell (RIAS A/S, Roskilde, Denmark) with an internal diameter of 8 cm, and compartment lengths of 5 cm and 10 cm, respectively, for the anolyte compartment and for the sample compartment. An anion exchange membrane, AR204SZRA, MKIII, Blank (Ionics, USA) separated the two compartments. The electrodes were Ti/MMO Permaskand wire with a diameter of 3 mm and a 50 mm length (Grønvold & Karnov A/S, Denmark). A power supply E3612A (Hewlett Packard, Palo Alto, USA) was used to keep a constant current in the cell. A magnetic stirrer at 250 rpm maintained the mixture in suspension into the sample compartment.

## 2.3 Methods

### 2.3.1 DES preparation and properties determination

DES were prepared by accurately weighing the amounts of the respective components in screw capped flasks, according to the appropriate molar ratios of each DES, in a Kern 770 weighing scale. The mixtures were then stirred and heated, until a clear homogeneous liquid was obtained (~24 h). When different percentages of distilled water were added, DES were also accurately weighed in order to obtain the desired water content (wt %).

The water content of the DES prepared was measured by Karl-Fischer (KF) titration, using an 831 KF Coulometer from Metrohm, with a generator electrode without diaphragm, and Hydranal Coulomat AG as reagent. The water content values obtained in weight percentage (wt %) are an average of at least three measurements for each DES.

DES viscosities and densities were determined using an SVM 3001 Viscometer (Anton Paar). The samples were submitted to a temperature scan, varying between 20 and 40 ºC, in 10 ºC intervals.

### 2.3.2 Experiments control, elements extraction and analysis

In all experiments, pH and conductivity were measured daily, respectively using a Radiometer pH-electrode EDGE (HANNA, USA) and a Radiometer Analytic LAQUA twin (HORIBA Ltd., Japan). Mining residues pH and conductivity were measured in a water suspension (L/S = 9). When an electric field was applied to the experiments, the voltage was also daily measured.

Total concentrations of As and W were determined by Inductively Coupled Plasma with Optical Emission Spectrometry (ICP-OES) (HORIBA Jobin-Yvon Ultima, Japan), equipped with generator RF (40.68 MHz), monochromator Czerny-Turner with 1.00 m (sequential), automatic sampler AS500 and dispositive Concomitant Metals Analyser. Also, initial As and W concentrations were determined by a TRACER 5 X-ray fluorescence equipment (XRF) (Bruker, Germany) considering a semiquantitive analysis. To quantify the elements in the solid matrix, aqua regia extraction was carried out mixing 0.5 g of mining residues with 3 mL of HCl (37 %) and 9 mL of $HNO_3$ (65 %). The vessels were set in a shaking table for 48 h at 140 rpm. Then, the samples were filtered by vacuum with 0.45 μm MFV3 glass microfibre filters (Filter lab, Barcelona, Spain) in order to separate any rest of solid present in the liquid to be further analyzed. Finally, the samples were diluted in a volumetric proportion of 1:25 with deionized water. Liquid samples (anolyte and liquid phase) were also filtered in the referred conditions, and all the samples were analyzed by ICP-OES in duplicates.

2.3.3 Data treatment and analysis

To determine the percentage of As and W that was removed (Eq. 1), separated (Eq. 2) and electromigrated (Eq. 3) in the ED process, the following equations were applied with the elements As/W (El) quantification data at the end of the experiments:

$$El\ removal\ (\%) = \frac{El_{liquid\ suspension}(mg) + El_{elecrolyte}(mg)}{El_{liquid\ suspension}(mg) + El_{elecrolyte}(mg) + El_{mining\ residues}(mg)} \times 100 \qquad (1)$$

$$El\ separation\ (\%) = \frac{El_{liquid\ suspension}(mg)}{El_{liquid\ suspension}(mg) + El_{elecrolyte}(mg) + El_{mining\ residues}(mg)} \times 100 \qquad (2)$$

$$El\ electromigration\ (\%) = \frac{El_{elecrolyte}(mg)}{El_{liquid\ suspension}(mg)} \times 100 \qquad (3)$$

The statistical data from all experiments were analyzed using the software GraphPad Prism version 7.0e. Statistically significant differences among samples for 95 % level of significance were calculated through ANOVA tests. The samples were compared according to the following criteria: (1) same element, same current, different DES; (2) same element, different current, same DES, and (3) different elements, same current, same DES.

3. RESULTS AND DISCUSSION

3.1 Matrix characterization

Mining residues were characterized using XRF and ICP-OES. The XRF results (Figure A in supplementary data) show that Si (silicon) represents 67.6 % of its total composition. The ICP results show the presence of high contents of As and W, 1675 and 130 mg/kg, respectively, but also of other elements, as Cu (71 mg/kg) and Fe (75150 mg/kg).

The high amount of Si is explained by the presence of quartz, among other silicate minerals in the ore (Candeias et al., 2014). The presence of silica is important once it can be turned into a pozzolanic/reactive material, increasing its durability and thus promoting a broad range of materials with application in the construction sector (Matias et al., 2014).

The pH of the mining residues, in water suspension, is slightly acid (5.3) and their conductivity is extremely low (0.8 mS/cm). Conductivity is an important variable for the experiments, where a current intensity was applied, in order to allow current passage and enable the electrolysis reactions. The addition of DES to all the experiments decreased the suspension pH to values below 2 and promoted sufficient ionic conductivity (up to 2.0 mS/cm) to maintain the current intensity applied during the selected period of 4 days (see section 3.2).

3.2 DES characterization and selection

DES have different water contents when compared to their individual initial components. A summary of the DES selected and their water contents is presented in Table 2, where it can be observed that water contents vary between 0.23 and 3 %. The data suggests that hydrogen bond network is maintained together with their properties as it is expected for DES with water contents bellow 40 or 50 wt. % (Dai et al., 2015; Hammond et al., 2017).

Density and viscosity are DES properties, directly related to their water content and composition that influence the efficiency of their applications. DES extremely viscous normally show lower extraction yields. Viscosity affect the solubilization/extraction yield of elements from solid matrices since mass transfer phenomena are dependent on DES viscosity.

Additionally, for the mining residues application, the DES pH´s also influence the solubility of the metal species and the extraction itself. DES containing carboxylic acids, as oxalic acid, malonic acid, lactic acid or propionic acid have an acidic pH (below 2), being more able to dissolve metals and metal oxides (Söldner et al., 2019). Concerning the carboxylic acids, the higher number of carboxyl and hydroxyl groups, as well as a shorter hydrocarbon chain, will result in higher DES viscosity values (Table A in supplementary data).

Preliminary tests were performed by adding four DES with different physico-chemical properties to mining residues, in order to screen their extraction efficiencies for As and W (Table 3). The best extraction yields were obtained with ChCl:MA for As (average: 16.2 %), and with ChCl:OA for W (average: 8.8 %). Table A (in supplementary data) presents the viscosities data for all experimental DES at room temperature, where viscosities of ChCl:OA and ChCl:MA were ~15,000 and ~4,000 mPa.s, respectively. Regarding W and As chemical properties, the extraction efficiency for each element is dependent on the used DES combination. DES composed by choline chloride may enhance W complexes release, since when high concentrations of chloride ions are available in solution, W may complex and form $[W_2Cl_9]^{3-}$.

The properties of the bidentate oxalate ion may have promoted the formation of complex oxalates with metals (Krishnamurty and Harris, 1961). Once W is in the form of $(Fe,Mn)WO_4$, oxalic acid may react with the solubilized $Fe^{2+}$, making $WO_4^{2-}$ available in solution.

Arsenic compounds may also be decomposed and stabilized by chloride ions, forming $AsCl_3$. Contrarily to W, a very stable element, As can be solubilized in extremely acidic conditions. On one hand, DES containing carboxylic acids, such as oxalic acid, malonic acid, lactic acid or propionic acid have an acidic pH (below 2). Comparing ChCl:OA, the most acid DES in the study, with ChCl:MA, the second has lower values of viscosity, which may have improved mass transfer phenomena in As case.

### 3.3 DES and electrical current combination

The combination of a current intensity (50 or 100 mA) with DES to enhance As and W extraction from the sample was tested. ChCl:MA and ChCl:OA were the selected DES for experiments 5-8, since they previously show to be the most efficient in As and W extraction, respectively (section 3.2 data). For these experiments, a power supply was connected to a pair of electrodes, and the cathode and the anode ends were placed inside a beaker, in order to generate an electric field. The current promoted the alkalization of the media along the time, more evident with ChCl:OA (Figure 2).

The media alkalization was faster for both DES when a current intensity of 100 mA was applied, achieving ChCl:OA a pH of 9.0 ± 4.3 and ChCl:MA a pH of 5.1 ± 0.8 at the end of the experiments (Figure 2). Both ChCl:OA and ChCl:MA provided an extremely acidic pH (below 2) to the initial sample. However, during the experiments the pH tends to increase. ChCl:OA is more viscous than ChCl:MA. Also, pKa from OA (1.46) is higher than MA (2.8), which means OA is more acid than MA. Thus, OA may had been consumed to form complexes with other elements (e.g. Cu, Fe), losing its buffer capacity. The result was a faster/higher pH increase when compared to the pH behavior in the ChCl:MA case, that is faster with a current intensity of 100 mA. In some cases, the pH may increase when temperature rises (Skulcova et al., 2018). The electric field generated inside the beaker may also have contributed to increase the temperature and, consequently, the pH overtime.

On the other hand, DES increased mining residues' conductivity, which allowed the passage of the current during the whole experiment time (Figure B in supplementary data). The conductivity of ChCl:OA (6.7 ± 3.8 mS/cm) was higher than ChCl:MA (2.0 ± 0.3 mS/cm) at the beginning of the experiments. However, ChCl:MA showed a stable conductivity along the time, with a slight variation (~8%). Contrarily, ChCl:OA decreased drastically its conductivity after 24 h, faster at 100 mA (to 2.1 ± 0.4 mS/cm), maintaining it stable in the remaining time. Oxalic acid may have reacted with elements present in solution, such as Fe (Mashaly et al., 2004) and Cu (Royappa et al., 2016). Complexes may have been quickly formed, lowing ions (e.g. Cl) in solution and, consequently, decreasing the media conductivity in 24 h. Also, since ChCl:OA is more viscous (Table A in supplementary data), ions transport may have been hindered (Craveiro et al., 2016).

Table 4 presents As and W extraction percentages obtained at the end of experiments 5-8. The average extractions were higher in experiment 8, with ChCl:OA at 100 mA (~28 % for As and ~16 % for W). Comparing the elements extraction between experiments with electrical current and with no electrical current application (Figure 3), there is an extraction yield upgrade in almost all cases. The highest improvements achieved were ~22 % for As extraction (ChCl:OA 100 mA) and ~11 % for W extraction (ChCl:MA 100 mA).

The combination of DES and electrical current may help to accelerate tungsten dissolution, since its chemistry is characterized by slow reactions. Paratungstates are the most important species by the progressive acidification of normal tungstate solutions in the pH range of 5–9, while metatungstates are stable in the pH range from 2.0 to 4.0. According toNguyen and Lee (2016), $WO_4^{2-}$ is predominant when pH is higher than 9, which corroborated the data obtained with experiment 8 (ChCl:OA 100 mA).

### 3.4 ED separation of As and W

Experiments 9-12 were carried out in the two-compartment ED cell (Figure 1). When the ED process was tested without DES, under the same experimental design and current intensities, the extraction of W was <1%. Thus, DES and mining residues were placed in the cathode compartment to assess the potential of extracted As and W to electromigrate into the anolyte, where they can later be separated. Once more, ChCl:MA and ChCl:OA were the selected DES, for the reasons previously explained.

The pH of the anolyte decreased to ~ 2 (Figure E in supplementary data), which was expected due to the water electrolysis generating $H^+$ at the anode. At the cathode compartment, during the first 24 h for both DES at 100 mA, an abrupt increase of pH to ~ 12 occurred, suggesting $OH^-$ generation by the water electrolysis at the cathode dominated (Figure C in Supplementary data).

Table 5 presents the removal percentages of As and W (extracted from the matrix and determined according to Eq. 1 in section 2.3.3), and the separation percentages of As and W (extracted from the mining residues and that electromigrated into the anode compartment, determined according to Eq. 2 in section 2.3.3). The highest simultaneous removal of As (35.0 ± 4.8 %) and W (22.1 ± 1.8 %) was observed with ChCl:OA (experiment 12). However, regarding the separation percentages, for As the highest was registered in experiment 11, with ChCl:MA (21.0 ± 1.2 %), while for W was in experiment 12, with ChCl:OA also (7.2 ± 3.1 %). Arsenic can be solubilized under acid conditions. Although the pH of the sample compartment increased during the experimental time, the anion exchange membranes separating the anode compartment from the cathode compartment does not work as a perfect rectifier, showering but allowing the passage of H+ to the cathode compartment. Thus, this passage of H+ to the cathode compartment may have contributed for the extraction and separation of As when the pH of the sample compartment turned to basic.

Considering the analytes percentage extracted from the matrix (Table 5), the ratios of As and W that electromigrated into the anode compartment (anolyte) were determined according to Eq. 3 (in section 2.3.3) and presented in Table 6. The high migration percentages were obtained with ChCl:MA at 100 mA for As (82.3 ± 1.3 %) and 50 mA for W (76.5 ± 8.6 %). This supported the hypothesis that As and W were in anionic forms and were able to cross the anion exchange membrane in the ED cell and hence to electromigrate into anolyte. Also, since ChCl:MA is less viscous, it may have facilitated the movement of species between compartments, improving the migration ratios.

These results proved the feasibility of coupling DES and ED to isolate As and W from mining residues in the anode compartment, from where they can be recovered. At the same time, the treated mining residues in the cathode compartment may be further used in construction materials, replacing conventional raw materials with economic and environmental benefits. Although the final residue is still not free of pollutants, leaching phenomena of harmful elements may be avoided with the studied treatment since the elements can be encapsulated. Also, there are emerging innovative strategies able to immobilize the remain contents of toxic metals in construction materials production, such as alkali activation (Kiventerä et al., 2018).

4. CONCLUSIONS

The experimental results proved the feasibility of coupling DES and ED to recover tungsten (W) and arsenic (As) from mining residues.

The study assessed the use of different deep eutectic solvents for the extraction of W and As from Panasqueira mine secondary resources on the tested experimental conditions. Their extraction efficiency was DES dependent. DES with choline chloride presented the higher average extraction yields, being ChCl:MA more selective for As (16 %), and ChCl:OA for W (9 %).

The study on current intensity influence with DES on As and W extractions (4 days experiments) showed an increasing in As and W extraction efficiencies, when compared to the tests with no current. The best average extractions were registered with ChCl:OA at 100 mA (28 % for As and 16 % for W), although with no statistically significant differences ($p < 0.05$).

A proof of concept of coupling DES with a two-compartment electrodialytic set-up was accomplished. The use of ChCl:OA at 100 mA for 4 days extracted simultaneously ~35 % of As and ~22 % of W from the matrix. The conditions generated in the cell sample compartment promoted the extraction and migration of As and W from the mining residues. A maximum of ~82 % of As and ~77 % of W extracted from the original matrix successfully electromigrated into the anode compartment, from where they can be further separated. The data support the applicability of electroremediation processes for this purpose.

Summing up, the experimental data suggests new possibilities for the recovery of critical raw materials and the removal of harmful compounds and elements from secondary mine resources that will contribute to circular economy and to increase sustainability in both mining and construction sectors that are seeking for innovative alternatives for materials production.

ACKNOWLEDGMENTS

This work has received funding from the European Union's Horizon 2020 research and innovation program under the Marie Skłodowska-Curie grant agreement No. 778045, as well as from Portuguese funds from FCT/MCTES through grant UID/AMB/04085/2019 and by the Associate Laboratory for Green Chemistry- LAQV UID/QUI/50006/2019. J. Almeida and A. Paiva acknowledge *Fundação para a Ciência e a Tecnologia* for, respectively, her PhD fellowship PD\BD\135170\2017, and for the project IF/01146/2015 attributed within the 2015 FCT researcher program. The authors acknowledge Carla Rodrigues from REQUIMTE for the ICP-OES analysis. This research is anchored by the RESOLUTION LAB, an infrastructure at NOVA School of Science and Technology.

**Figure captions**

Figure 1 - Electrodialytic two-compartment cell, with stirrer and anion exchange membrane (AEM), connected to a power supply

Figure 2 - pH behavior in the cathode compartment during experiments 5-8, corresponding to DES with an applied current intensity

Figure 3 - Improvement percentages of arsenic and tungsten extraction with a current intensity of 50 mA (experiments 5, 6) and 100 mA (experiments 7, 8) compared to the experiments with no current (experiments 1, 2)

**Table legends**

Table 1 - Experimental conditions: with DES extraction; DES and current intensity extraction; and DES and electrodialytic separation

Table 2 - Identification of the selected DES, molar ratio, designation, water content and molecular structure

Table 3 - Arsenic and tungsten extraction percentages from mining residues obtained with different DES

Table 4 - Arsenic and tungsten extraction percentages with ChCl:MA and ChCl:OA, at 50 and 100 mA

Table 5 - Removal and separation percentages of As and W obtained at the end of experiments coupling DES with ED

Table 6 - Percentages of extracted As and W that electromigrated to the anolyte

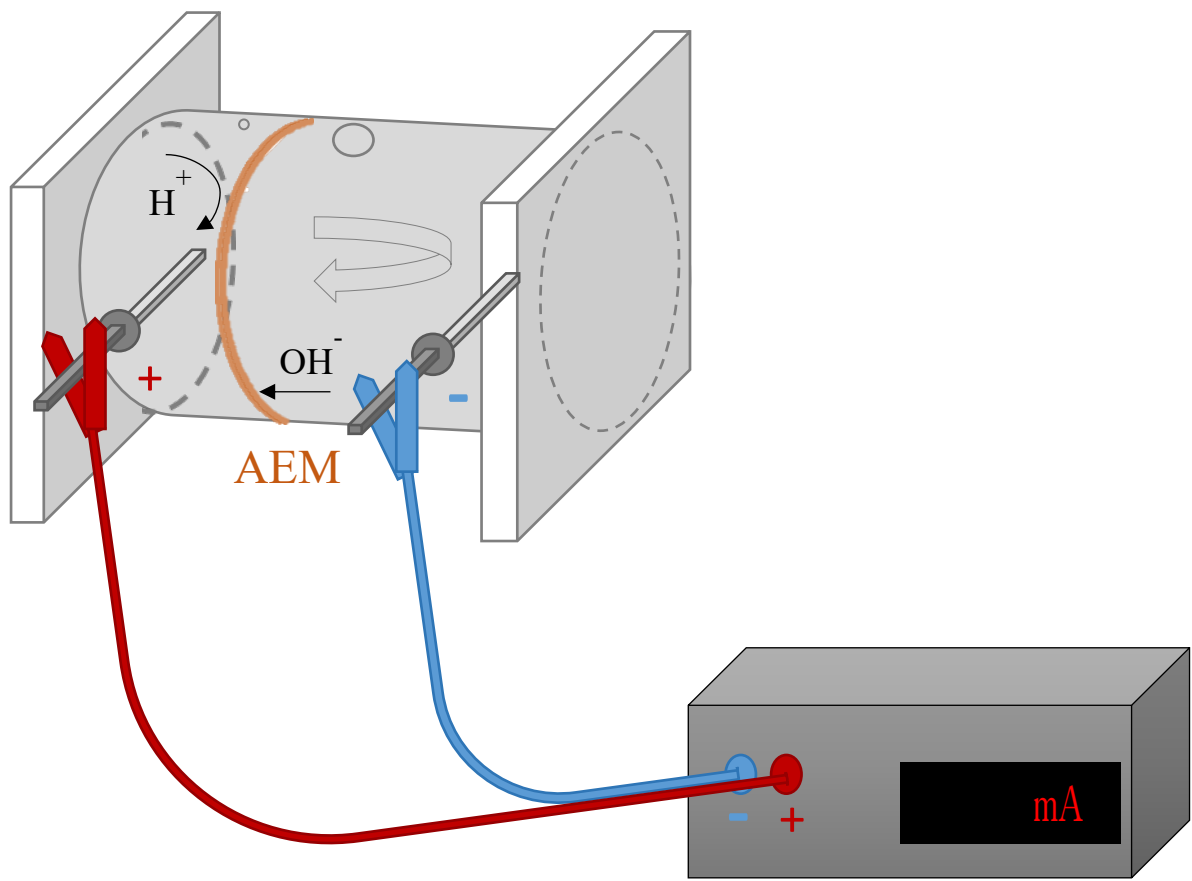

Figure 1 - Electrodialytic two-compartment cell, with stirrer and anion exchange membrane (AEM), connected to a power supply

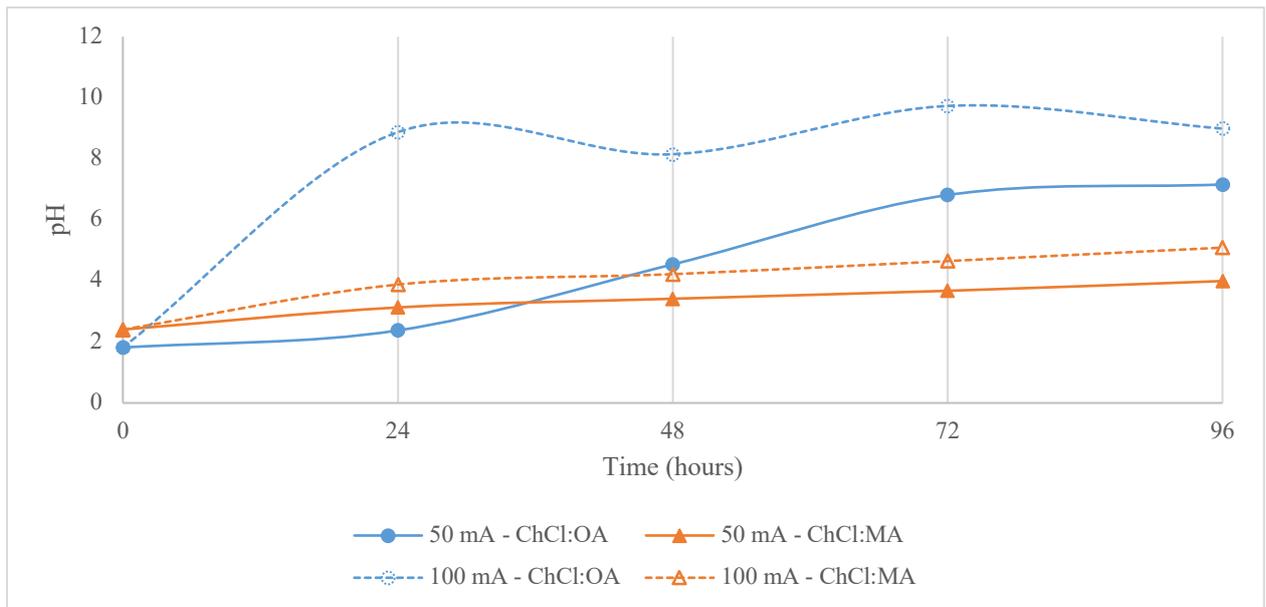

Figure 2 –pH behavior in the cathode compartment during experiments 5-8, corresponding to DES with an applied current intensity

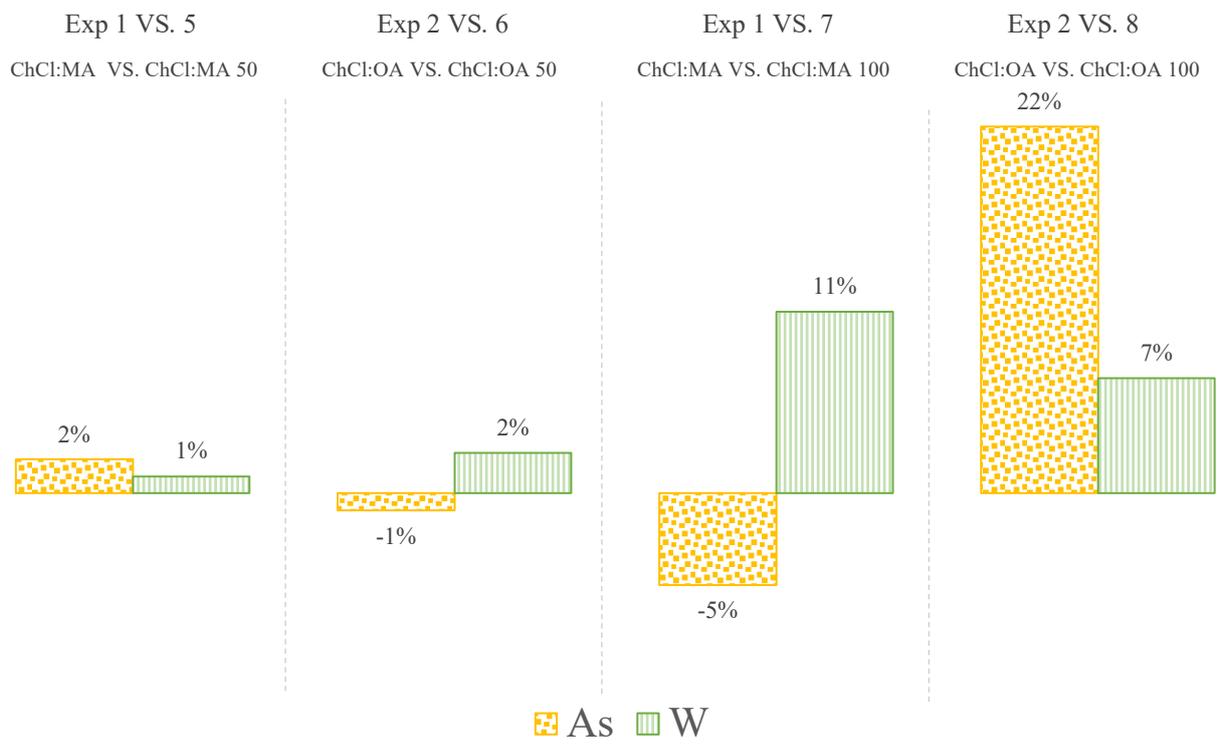

Figure 3 – Improvement percentages of arsenic and tungsten extraction with a current intensity of 50 mA (experiments 5, 6) and 100 mA (experiments 7, 8) compared to the experiments with no current (experiments 1, 2).

Table 1 - Experimental conditions: with DES extraction; DES and current intensity extraction; and DES and electrodialytic separation

| Experiment | Code | Liquid solution proportions | Duration (days) | Current intensity (mA) | Set-up |
|---|---|---|---|---|---|
| DES extraction | | | | | |
| 1 | ChCl:MA | 100 % Choline Chloride:Malonic Acid (1:2) | 10 | - | |
| 2 | ChCl:OA | 100 % Choline Chloride:Oxalic Acid (1:1) | 10 | - | 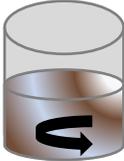 |
| 3 | ChCl:LA | 100 % Choline Chloride:Latic Acid (1:2) | 10 | - | |
| 4 | PA:U | 100 % Propionic Acid:Urea (2:1) | 10 | - | |
| DES + electrical current | | | | | |
| 5 | ChCl:MA 50 | 5 % Choline Chloride:Malonic Acid (1:2) + 95 % Deionized water | 4 | 50 | |
| 6 | ChCl:OA 50 | 5 % Choline Chloride:Oxalic Acid (1:1) + 95 % Deionized water | 4 | 50 | 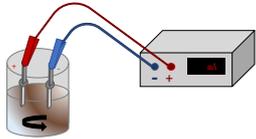 |
| 7 | ChCl:MA 100 | 5 % Choline Chloride:Malonic Acid (1:2)+ 95 % Deionized water | 4 | 100 | |
| 8 | ChCl:OA 100 | 5 % Choline Chloride:Oxalic Acid (1:1) + 95 % Deionized water | 4 | 100 | |
| DES + ED | | | | | |
| 9 | ChCl:MA 50 ED | 5 % Choline Chloride:Malonic Acid (1:2) + 95 % Deionized water | 4 | 50 | |
| 10 | ChCl:OA 50 ED | 5 % Choline Chloride:Oxalic Acid (1:1) + 95 % Deionized water | 4 | 50 | 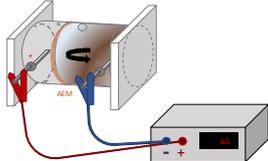 |
| 11 | ChCl:MA 100 ED | 5 % Choline Chloride:Malonic Acid (1:2) + 95 % Deionized water | 4 | 100 | |
| 12 | ChCl:OA 100 ED | 5 % Choline Chloride:Oxalic Acid (1:1) + 95 % Deionized water | 4 | 100 | |

Table 2 - Identification of the selected DES, molar ratio, designation, water content and molecular structure

| DES | Molar ratio | Designation | Water content (wt. %) | Molecular structure |
|---|---|---|---|---|
| Choline Chloride:Malonic Acid | (1:2) | ChCl:MA (1:2) | 3.0 | |
| Choline Chloride:Oxalic Acid | (1:1) | ChCl:OA (1:1) | 0.23 | |
| Choline Chloride:Lactic Acid | (1:2) | ChCl:LA (1:2) | 0.74 | |
| Propionic Acid:Urea | (2:1) | PA:U (2:1) | 0.78 | |

Table 3 - Arsenic and tungsten extraction percentages from mining residues obtained with different DES

| Experiment | Code | As extraction (%) | W extraction (%) |
|---|---|---|---|
| 1 | ChCl:MA | 16.2 ± 5.9[A] | 4.4 ± 0.8[a] |
| 2 | ChCl:OA | 6.0 ± 0.9[a] | 8.8 ± 0.3 |
| 3 | ChCl:LA | 4.0 ± 2.4[a] | 0.9 ± 0.1 |
| 4 | PA:U | 7.7 ± 1.7 | 5.5 ± 0.9 |

*Statistical analysis: Multiple comparisons were statistically performed at p<0.05 (95 % confidence interval); data with capital letter is statistically significantly different from the data with the same lowercase letter.*

Table 4 - Arsenic and tungsten extraction percentages with ChCl:MA and ChCl:OA, at 50 and 100 mA

| Experiment | Code | As extraction (%) | W extraction (%) |
|---|---|---|---|
| 5 | ChCl:MA 50 | 18.2 ± 15.8 | 5.4 ± 2.8 |
| 6 | ChCl:OA 50 | 5.0 ± 5.9 | 11.2 ± 9.5 |
| 7 | ChCl:MA 100 | 10.6 ± 10.2 | 15.1 ± 9.6 |
| 8 | ChCl:OA 100 | 27.6 ± 26.3 | 15.6 ± 11.5 |

*There are no statistically significantly differences in the results presented at p<0.05 (95 % confidence interval).*

Table 5 - Removal and separation percentages of As and W obtained at the end of experiments coupling DES with ED

| Experiment | Code | As (%) Removal/Separation | W (%) Removal/Separation |
|---|---|---|---|
| 9 | ChCl:MA 50 ED | 10.8 ± 1.6[b] / 4.9 ± 0.5[e] | 1.6 ± 0.38 / 0.6 ± 0.1 |
| 10 | ChCl:OA 50 ED | 9.2 ± 2.7[a] / 5.6 ± 2.8[f] | 7.4 ± 3.7[c] / 5.4 ± 2.2 |
| 11 | ChCl:MA 100 ED | 25.4 ± 1.1[B,d] / 21.0 ± 1.2[E,h] | 7.2 ± 1.2[D] / 4.3 ± 0.1[H] |
| 12 | ChCl:OA 100 ED | 35.0 ± 4.8[A,c] / 17.0 ± 4.6[F,g] | 22.1 ± 1.8[C,d] / 7.2 ± 3.1[G] |

*Statistical analysis: Multiple comparisons were statistically performed at $p<0.05$ (95 % confidence interval); data with capital letter is statistically significantly different from the data with the same lowercase letter.*

SUPPLEMENTARY DATA

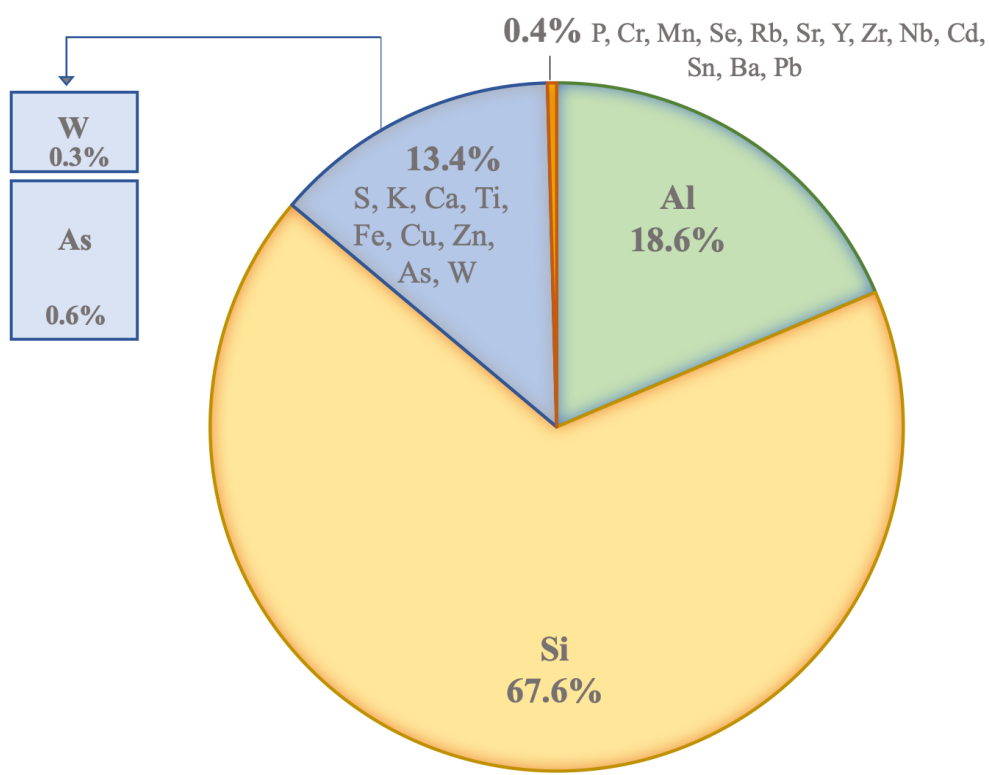

Figure A - Elements distribution in the mining residues obtained by X-ray fluorescence (Semiquantitative data, wt. %)

Table A – Deep eutectic solvents dynamic viscosity between 20 and 40 ºC (mPa.s)

| Temperature (ºC) | mPa.s | | | |
| --- | --- | --- | --- | --- |
| | ChCl:MA | ChCl:OA | ChCl:LA | PA:U |
| 20 | 3,891 | 14,683 | 257 | 22 |
| 30 | 1,251 | 5,020 | 137 | 14 |
| 40 | 538 | 2,007 | 78 | 9 |

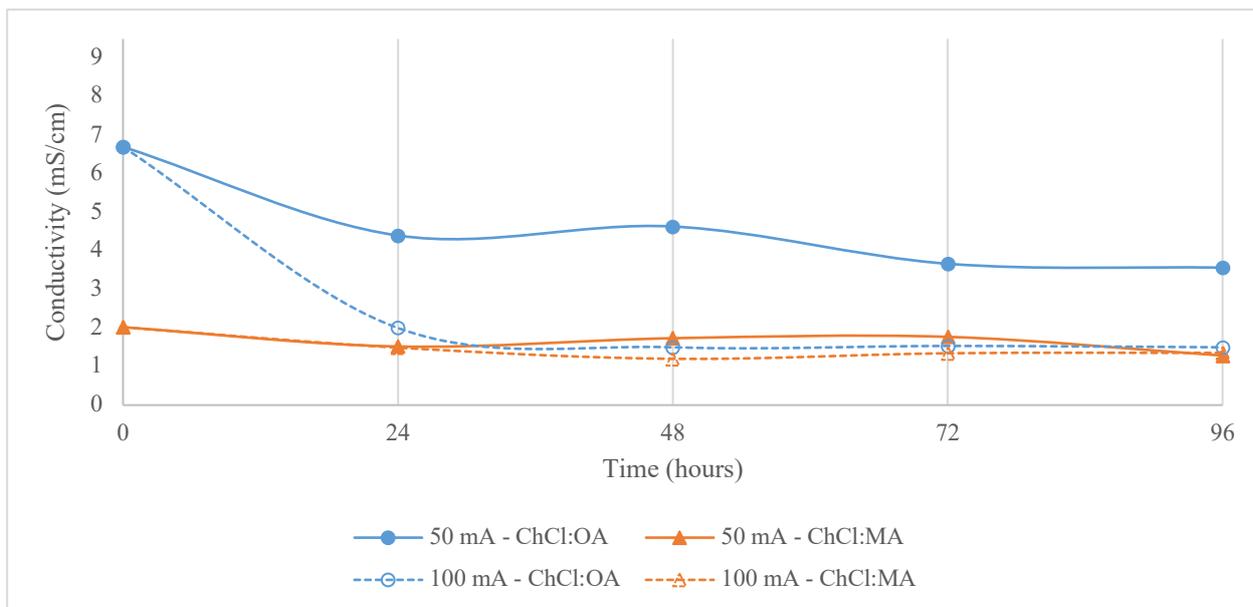

Figure B - Conductivity behavior during experiments 5-9, corresponding of DES with an applied current intensity

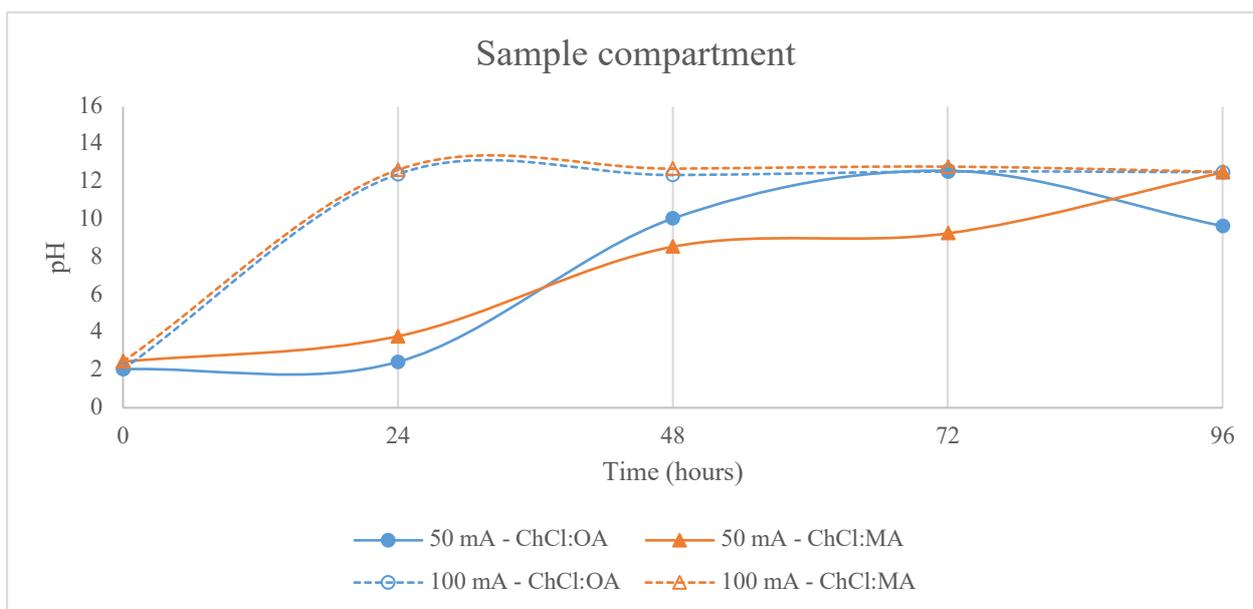

Figure C - pH measured at the cathode compartment of the ED set-up during the experiments 9-12, corresponding to DES + ED.

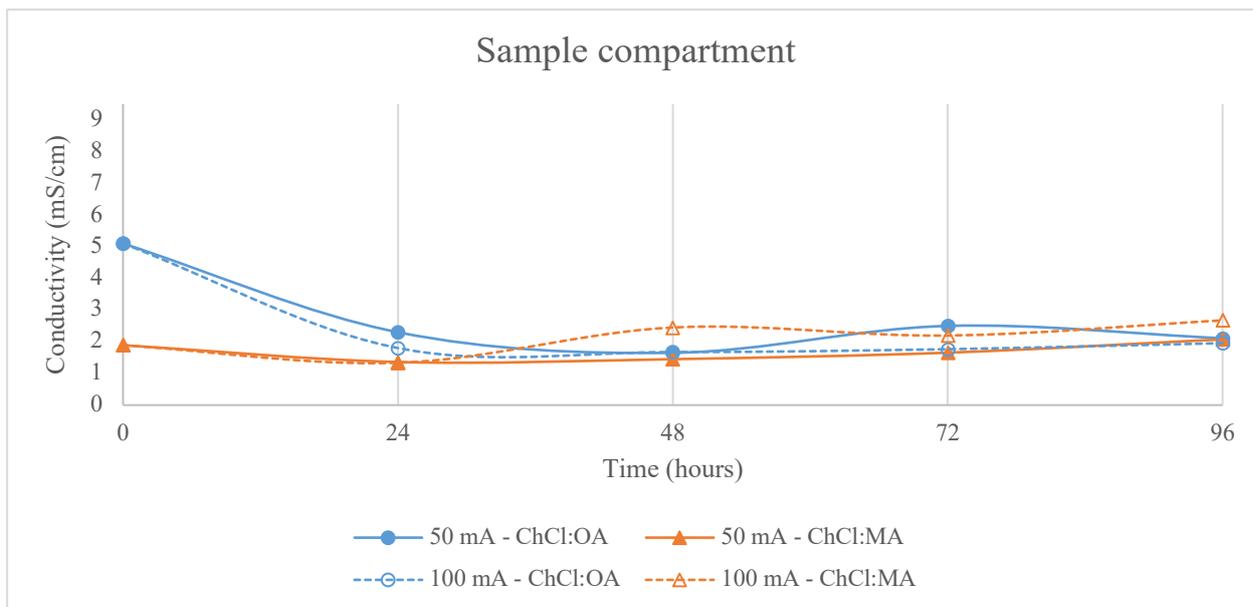

Figure D - Conductivity at the cathode compartment of the ED set-up during the experiments 9-12, corresponding to DES + ED.

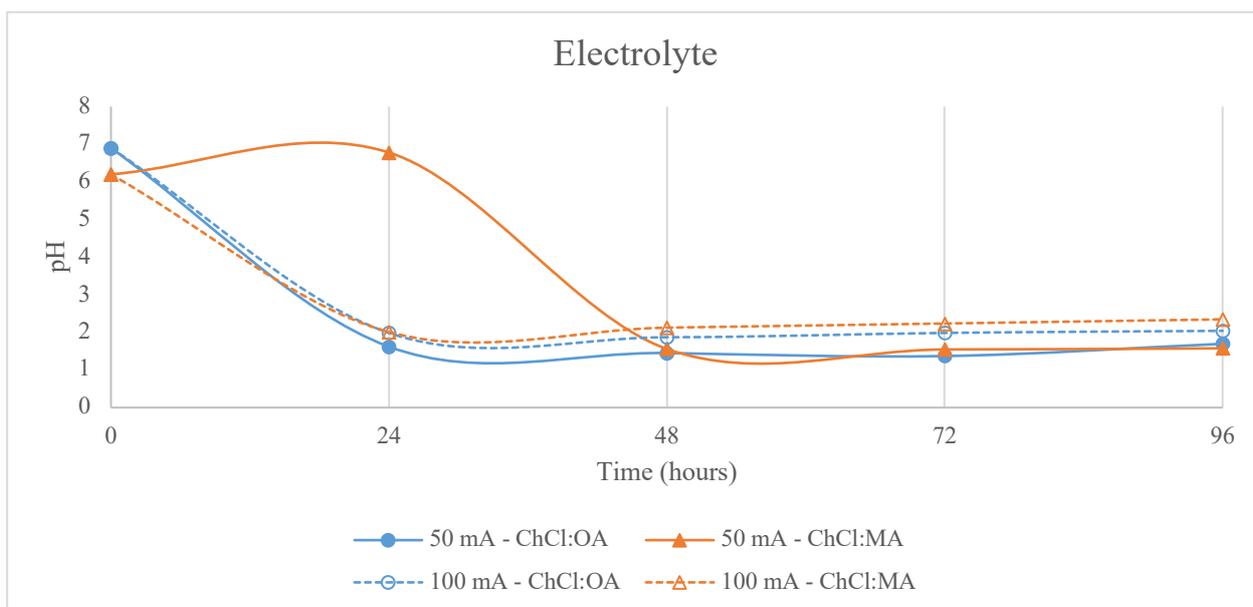

Figure E - pH measured at the anode compartment of the ED set-up during the experiments 9-12, corresponding to DES + ED.

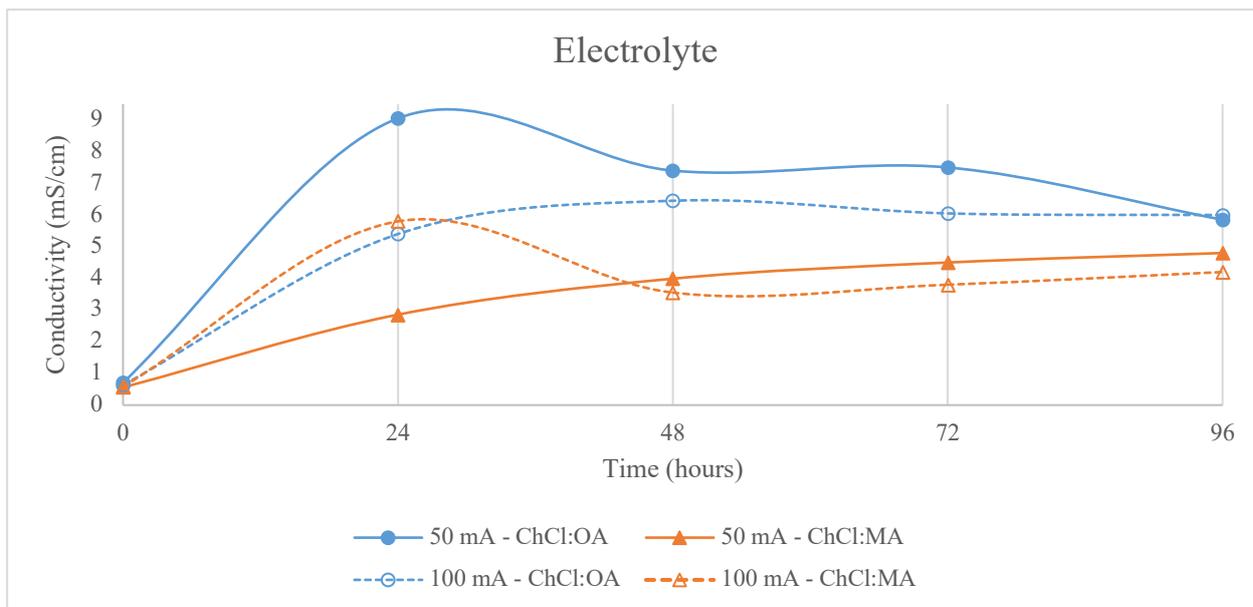

Figure F - Conductivity at the cathode compartment of the ED set-up during the experiments 9-12, corresponding to DES + ED.